\documentclass[twocolumn]{aa}
\input colordvi.sty
\usepackage{graphicx}
\usepackage{txfonts}
\begin{document}
\title{3-D numerical simulations of rotating jets}

\subtitle{The case of the DG Tau microjet}

\author{A.H. Cerqueira
       \inst{1}
       \and
       E.M. de Gouveia Dal Pino\inst{2}
       }

\offprints{A.H. Cerqueira}

\institute{LATO-DCET-UESC, Rodovia Ilh\'eus-
Itabuna, km 16, Ilh\'eus, Bahia, Brazil, CEP 
45662-260\\
\email{hoth@uesc.br}
\and
IAG-USP, Rua do Mat\~ao, 1226, Cidade 
Universit\'aria, S\~ao Paulo, S\~ao Paulo, Brazil, 
CEP: 01060-970\\
\email{dalpino@astro.iag.usp.br}
           }

%   \date{Received ; accepted }

\abstract{

We here present results of three-dimensional Smoothed Particle hydro and
magnetohydrodynamics  simulations of rotating jets, also including the
effects of radiative cooling, precession and velocity variability. Using
initial conditions and parameters which are particularly suitable for the
DG Tau microjet, we have been able to approximately reproduce its complex
knotty morphology and kinematics.  We have also obtained radial velocity
maps which are in good agreement with the data obtained by Bacciotti et al., 
thus indicating that their interpretation that the DG Tau
microjet is rotating is correct. Finally, we have found that a magnetic
field of the order of $\approx 0.5$ mG is sufficient to collimate the jet
against the lateral expansion that is caused by the centrifugal forces.

        }

\maketitle
%
%______________________________________

\section{Introduction}

The seek for the real nature of the accretion-ejection mechanism
underlying the formation of the jets is one of the major contributions of
the detailed analysis of Herbig-Haro (HH) jet data.  The statement that
MHD models can be responsible for generating the jets associated with
Young Stellar Objects is still a matter of debate.  The theoretical
discussion of the so-called magneto-centrifugal models for jet
launching (e.g., Blandford \& Payne \cite{bland82}, K\"onigl 1982,
Spruit \cite{spruit96}), can be strongly constrained by the existence of
observational data.

Recent observations of microjets associated with T-Tauri stars, namely,
DG Tau, RW Aur, TH28 and LkH$\alpha$321 (Bacciotti et al \cite{bacci02},
Coffey et al \cite{coffey04}), have revealed important trends in the
jet radial velocity field that are consistent with a rotation pattern
inside the jet beam. The presence of  rotation could be understood in the
context of the magneto-centrifugal models as due to  jet launching from
a Keplerian accretion disk along with the magnetic field lines which
are anchored into the disk-star system and provide the collimation of
the beam.

Motivated by these recent findings, we have carried out a set of
fully three-dimensional  simulations of rotating jets, also taking into
account the presence of radiative cooling, jet precession, and  velocity
variability (to allow the production of internal knots). All these are
necessary  ingredients to address the kinematical complexity of the
microjets and we are here particularly interested in the DG Tau microjet
case, which is one of the best studied  in the literature. We have also
investigated, in few models, the collimating effects of magnetic fields
upon rotating jets.

\section{Numerical method and the simulated models} 

In order to examine the behaviour of the microjets and their early
evolution near the source under the effects of both rotation and
precession, we here employ modified versions of our three-dimensional
Smoothed Particle Hydrodynamics  (3-D SPH) and Magnetohydrodynamics (3-D
SPMHD) codes which were previously designed to investigate the large scale
structure and evolution of protostellar jets either in the absence (see,
e.g., de Gouveia Dal Pino 2001) or in the presence of magnetic fields
(e.g, Cerqueira \& de Gouveia Dal Pino 2001).

\begin{table}
\caption[1]{The jet models}
{\small
\centerline{
\begin{tabular}{|c|c|c|c|c|}
\hline\hline
\noalign{\smallskip}
\hbox{Model} & $\tau_{\rm pul}$ & $\tau_{\rm prec}$ 
& rotation? & B-field? \\
\noalign{\smallskip}
\hline
\noalign{\smallskip}
\hbox{A} & \hbox{-} & \hbox{8 years} & 
\hbox{no} & \hbox{no} \\
\hbox{B} & \hbox{8 years} & \hbox{8 years} & 
\hbox{no} & \hbox{no} \\
\hbox{C} & \hbox{8 years} & \hbox{8 years} & 
\hbox{yes} & \hbox{no} \\
\hbox{D} & \hbox{-} & \hbox{-} & \hbox{yes} 
& \hbox{no} \\
\hbox{E} & \hbox{-} & \hbox{-} & \hbox{yes} 
& \hbox{yes} \\
\noalign{\smallskip}
\hline
\noalign{\smallskip}
\end{tabular}
}
}
\end{table}

Our computational domain is a rectangular box that mimics the ambient
medium and has dimensions $-25 R_{\rm j} \le$ x $\le 25 R_{\rm j}$, 
and $-10R_{\rm j}
\le$ y,z $\le 10 R_{\rm j}$, where $R_{\rm j}$ is the initial jet radius (which
is also the code distance unit). The Cartesian coordinate system has
its origin at the  center of the box and the jet flow is continuously
injected into the bottom of the box [at the inlet which has coordinates
${\bf r}=(-25R_{\rm j},0,0)$] with  top-hat density and pressure profiles.
The jet and gas pressures are assumed to be in equilibrium at the inlet.
Inside the box, the SPH particles are initially regularly distributed
in a cubic lattice. An outflow boundary condition is assumed for the
boundaries of the box. The particles are smoothed out by a spherically
symmetric kernel function of width $h$, and the initial values of $h$ were
chosen to be $0.4 R_{\rm j}$ and $0.2R_{\rm j}$ for the ambient and jet particles,
respectively. The adiabatic index of the ambient medium and the jet is
assumed to be $\gamma=5/3$, and an ideal equation of state is used. The
radiative cooling, due to collisional excitation and recombination, is
implicitly calculated using a time-independent cooling function for a gas
of cosmic abundances cooling from $T \simeq 10^6$ to $9 \times 10^3$ K.

\begin{figure}
\centering
\includegraphics[width=8.5cm]{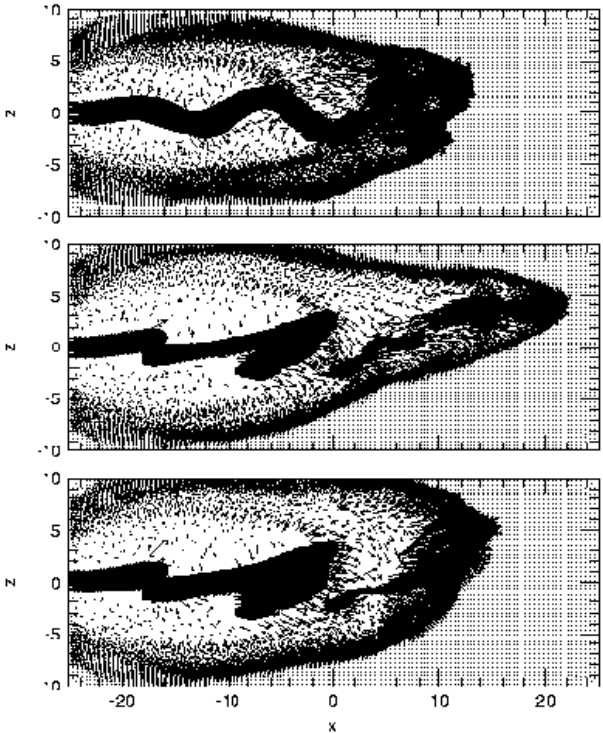}
\caption{Midplane velocity field distribution for 
the jet models
A (top), B (middle) and C (bottom), at 
$t/t_{\rm d}=2.5$. See the text
for details.}
\label{FigVibStab}
\end{figure}

The adopted parameters for the simulations are appropriate to the
conditions generally found in protostellar jets and, in particular,
in the DG Tau microjet.  We take a number density ratio between the
jet and the ambient medium $\eta=n_{\rm j}/n_{\rm a}=10$; $n_{\rm j}=2000$ 
cm$^{-3}$;
an average ambient Mach number $M_{\rm a} = v_{\rm j}/c_{\rm a}=19$, 
where $v_{\rm j} \simeq $
300 km s$^{- 1}$ is the average jet velocity of DG Tau (e.g., Bacciotti
et al. 2002, Pyo, Kobayashi, Hayashi et al.  2003), and $c_{\rm a} \simeq 16 $ km s$^{-1}$ is the
ambient sound speed for a gas with an average temperature of  $10^4$ K;
and $R_{\rm j} = 5.6 \times 10^{14}$ cm (see e.g., Dougados et al. \cite{dou00},
Bacciotti et al. 2002).

We have carried out several simulations considering both continuous
and pulsing jets (see below). In the latter cases, we have  adopted a
sinusoidal profile to describe the ejection velocity time- variability
at the inlet: $v_{\rm o}(t)= v_{\rm j} [1 + A\cdot{\rm sin}({{2 \pi}\over 
\tau_{\rm pul}} t)]$,

\noindent  where $A$ is the velocity amplitude and $\tau_{\rm pul}$ is
the period of the oscillation.  The values for these parameters were
obtained from observations of the DG Tau microjet (see, e.g., Lavalley
et al. 1997, Raga et al. \cite{raga01}) which suggest $A=0.33$, and $
\tau_{\rm pul} = $ 8 years = $0.71 t_d$ (where $t_{\rm d}=R_{\rm j}/c_{\rm a} 
\approx 11.3$
years corresponds to the transverse jet dynamical time).  In the numerical
models where the jet precession has been also taken into account, we have
assumed an equal precession period (i.e., $\tau_{\rm prec}=\tau_{\rm pul}=8$
years) and a precession half-angle $\theta = 5^o$ as indicated by the
DG Tau jet observations (Lavalley-Fouquet et al. 2000, Dougados  et al.
\cite{dou00}).

For the computation of the jet rotation around its main axis, we have
assumed that the flow conserves its angular momentum after its passage
through the Alv\'en surface (i.e., $v_{\phi}\cdot r =$ constant, where
$v_{\phi} $ is the toroidal velocity in the normal direction to the jet
axis). This is consistent with magneto-centrifugal mechanisms for jet
launching (see, e.g., Spruit \cite{spruit96}) and is also suggested by
the observations of the DG Tau microjet (Bacciotti et al. \cite{bacci02})
which imply $v_{\phi} = (4.362 \times 10^{20}/r)~{\rm cm}~{\rm s}^{-1}$
where  $r$ is the radial distance to the jet axis in cm. This gives
$v_{\phi} \simeq 8$ km s$^{-1}$ at the jet surface ($r=R_{\rm j}=5.6\times
10^{14}$ cm), and  $v_{\phi} \approx 55$ km s$^{-1}$ near the jet axis
(at $r \simeq 0.15R_{\rm j}$) \footnote{We notice that the code does not
allow, by construction, any SPH particle to occupy the jet axis, so
that the equation above for $v_{\phi}$ has no singularities, as it is
employed only for values of $r$ which are larger than zero.}. It should be
emphasized that we are here mostly interested in carrying out simulations
of jet regions near the source but beyond the Alfv\'en surface (e.g.,
Spruit \cite{spruit96})  where the flow has accelerated to velocities
larger than the Alfv\'en speed and is believed to be already collimated,
and which are, at least  in some cases, partially resolved by optical
observations (as for instance, DG Tau).

\begin{figure}
\centering
\includegraphics[width=8.5cm]{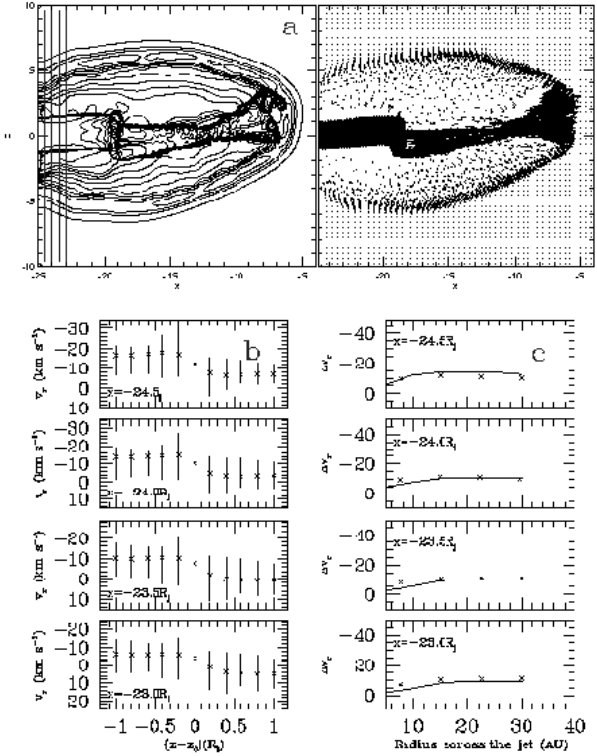}
\caption{ (a) Midplane density contour (left panel) 
and velocity
field (rigth panel) distributions for the model C 
(see Table 1),
at $t/t_{\rm d}=1$ which approximately corresponds  to 
the dynamical age
of the DG Tau microjet; (b) Average radial 
velocities  across the jet  for the
different positions along the beam that are 
marked in Figure 2a, above. We notice that the 
jet axis is actually located at $z_{\rm o} = 0.1 R_{\rm j}$ due 
to the effects of the precession. The vertical bars 
indicate the maximum and minimum values of 
the radial velocities obtained for  each radial 
distance
(the distances are in units of  the jet radius); (c) 
Radial velocity differences between positions 
across the jet which are symmetric with respect to 
the jet center, 
as a function of  the radial distance. Crosses: our 
results. Full line:
an approximate fitting to Bacciotti et al. (2002) 
observations and  Pesenti, Dougados, Cabrit et al (\cite{pesenti}) 
results (the distances are in units of
AU and the velocities in units of  km s$^{-1}$).} 
\label{FigVibStab} 
\end{figure}

Table 1 summarizes the properties of the simulated models. In model A,
we have run a precessing jet with constant velocity at injection and
no rotation. In model B, we have considered a $pulsing$ precessing
jet also without rotation, and in model C we have run a jet which is
simultaneously pulsating, precessing, and rotating. All these three
models are purely hydrodynamical and share the same initial conditions
as described above. Their velocity distribution is compared in Figure 1
after they have propagated over a time $t/t_{\rm d}=2.5$. In model C, the jet is
precessing counter clockwise while rotating clockwise. The jet of model
A develops a V-shape structure at the head which is commonly detected
in precessing jets (e.g., Masciadri et al.  \cite{mas02}, Masciadri \&
Raga \cite{mas03}, Rosen \& Smith \cite{rosen04}) and is due to the impact
of the parcels propagating in different directions in the wiggling flow
with the bow-shock front. In the jet of model B, the combined effects of
pulsation and precession produce internal working surfaces propagating
in different directions. In this case, a narrower bowshock structure
develops at the head that makes it to  propagate slightly faster than the
jet of model A. The rotating jet of model C essentially bears the same
features of the previous models but develops an even larger head than
model A due to the centrifugal forces that tend to push the jet flow
outwards in the radial direction (see also Figure 3 and discussion below).

Let us now concentrate on the early evolution of the rotating, pulsing,
precessing jet of Figure 1c (model C) whose characteristics qualitatively
resemble those of the DG Tau microjet (Dougados et al.  \cite{dou00},
Bacciotti et al. \cite{bacci02}).  Figure 2a shows the jet of Figure 1c
at an earlier time that corresponds to the inferred dynamical age of the
DG Tau microjet ($\sim$ 11.3 yr; see e.g., Dougados et al \cite{dou00};
Lavalley et al \cite{lava}; Pyo, Kobayashi, Hayashi et al. 2003). 
The density contour plot
presents nearly similar features to those observed in the DG Tau microjet:
an elongated, highly collimated beam near the jet base, a slight sideway
displacement of the  beam that has been previously interpreted as an
evidence for precession of the jet axis, and a prominent bow shock (at
$\approx -8R_{\rm j}$ in Figure 2a) which can be compared with the B1 knot of DG
Tau (Dougados et al.  \cite{dou00}, Lavalley-Fouquet et al. \cite{lava2}).

For a comparison with the radial velocities obtained from the HST
observations of DG Tau by Bacciotti et al.  (\cite{bacci02}), in
Figure 2b we have built diagrams of  the radial velocity across the
jet (taken at an angle of $45^{\circ}$ with respect to the jet axis,
in order to mimic the DG Tau inclination with respect to the line of
sight) in four different slices along the flow which are separated by $
0.5R_{\rm j} \simeq 20$ AU and are marked in Figure 2a.  [These correspond   to
the same slices examined in Figure 1 of Bacciotti et al.: $x=-24.5R_{\rm j}$
(Position I), -24$R_{\rm j}$ (Position II), -23.5$R_{\rm j}$ (Position III) and
-23.0$R_{\rm j}$ (Position IV).] Figure 2c depicts, for each diagram of
Figure 2b, the radial velocity differences between positions which are
symmetric with respect to the central region of the jet, as a function
of the distance from the jet axis.  We notice that the differences
in radial velocity are negative, therefore indicating that the net
velocity is oriented towards the $-y$ direction. These negative shifts
in velocity are obviously a consequence of the fact that the jet of
Model C is rotating clockwise.  As expected, these shifts are null in
the jets that are only precessing, like those of Figures 1a and b. The
resulting radial velocity shifts in Figure 2c, $\Delta v_{\rm r} \lesssim $
-15 km s$^{-1}$, are very similar to those obtained by Bacciotti et al.
which are approximately reproduced in the figure by the solid lines.
We note that these solid lines also reproduce  the results recently
obtained by Pesenti, Dougados, Cabrit et al.  
(\cite{pesenti}) assuming that the DG Tau
microjet is launched magnetocentrifugally by a warm accretion disk.
The consistency between the present results and their predictions is an
additional support for the magnetocentrifugal mechanism for jet launching.

\begin{figure}
\centering
\includegraphics[width=8cm]{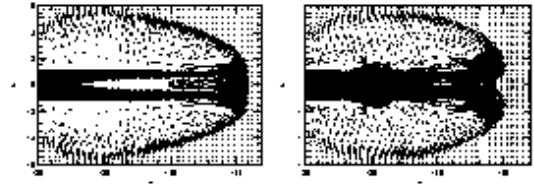}
\caption{Midplane velocity field distributions for 
the purely
hydrodynamical model D (left) and for the MHD 
model E (right) of Table 1,
at $t/t_{\rm d}=1$.  The distances are in units of $R_{\rm j}$ 
($\approx 38$ AU).}
\label{FigVibStab}
\end{figure}

Finally, aiming to check the collimation  effects of the magnetic
fields in rotating jets, we have performed simulations  of continuous,
non-precessing flows under the presence of helical fields.  Given the
present uncertainties related to the real orientation and strength of
the magnetic fields in protostellar jets we have assumed, as in previous
work (e.g., Cerqueira \& de Gouveia Dal Pino 2001) a force-free helical
magnetic field which also extends to the ambient medium and whose radial
functional dependences for the toroidal and longitudinal components are
respectively given by:

$$B_{{\phi}}(r)=B_0 [f(r) d]^{1/2},
\eqno(1a)$$
$$B_x(r)=B_0 [1 -f(r) (r+d)]^{1/2},
\eqno(1b)$$
 
\noindent where 
$r=\sqrt{y^2+z^2}$ is the radial 
distance from the
jet axis, 
$f(r) = 0.5 C r^2/ (r+0.5 d)^3$,
and the constants $C$ and $d$ are 
assumed to be  100
and $3R_{\rm j}$, respectively. In these equations, 
$B_0$ gives the maximum
strength of the magnetic field and corresponds to 
the magnitude
of  the longitudinal component at the jet axis and 
the magnitude of the
toroidal component at $r=3R_{\rm j}$ (see Cerqueira 
\& de Gouveia Dal Pino 2001). 

As an example, Figure 3 compares the early evolution of one of these
MHD runs (model E of Table 1) with a purely  hydrodynamical jet (model
D of Table 1). Both have the same initial conditions of the previous
models which are appropriate to DG Tau microjet. The MHD model has an
initial value of the ratio between the magnetic and the gas pressure
$\beta =p_{\rm mag}/p_{\rm th}=B^2/(8\pi p) = $10 near the jet surface which
corresponds to a magnetic field intensity $\sim 0.5$ mG.  The centrifugal
force upon the fluid in the radial directions causes the beam radius
of the  hydrodynamical jet to increase as the central parts of the flow
tend to move laterally outwards.  Though less obvious, we have previously
noticed the same unbalancing of radial forces and jet enlargement in the
rotating jet of Figure 1c (Model C), particularly in its head. When a
helical magnetic field (with surface $\beta \simeq$ 10) is introduced in
the system, we find that the tension force (or hoop stress) associated to
the toroidal component of the magnetic field is able to collimate the jet.
Smaller values of the surface $\beta$ were found to be unable to prevent
the lateral expansion and similar results were also obtained for pulsing
jets under the same conditions (see discussion below).

\section{Discussion and conclusions}

In the present work, we have explored the role of rotation on the
evolution of  YSO microjets addressing, in particular, the DG Tau
microjet that has recently shown strong observational evidence for
rotation (Bacciotti et al. \cite{bacci02}). With the help of  3-D hydro
and magnetohydrodynamical simulations, we have investigated several
models that  also included the effects of precession and  pulsation
of the flow.  Our main purpose here was to verify through fully 3-D
simulations whether the signature of rotation was really  unique and
unmistakable in the jet flow when other  effects,  like precession and
pulsation, were also considered.  The main conclusions of this study
could be summarized as follows:

1. The morphology and kinematics of the DG Tau microjet was best
reproduced in our simulations by model C (Figure 1c and 2a), which
presents a jet with a sinusoidal velocity variability (with  mean velocity
of 300 km s$^{- 1}$, half-amplitude of 100 km s$^{-1}$, and a period of
8 yr), a precession of the outflow axis (with  half-angle of 5$^{\circ}$
and 8 yr period), and a rotation around its axis with a velocity $v_{\phi}
\simeq 8$ km s$^{-1}$ at the jet surface, and $\approx 55$ km s$^{-1}$
near the jet axis,. These latter conditions satisfy angular momentum
conservation and  are consistent with magneto-centrifugal jet launching
models.  As remarked before, these characteristics approximately match
those suggested by the observations reported in the literature.  We note
also that previous numerical studies of the morphology and kinematics of
this object (e.g., Raga et al. \cite{raga01})  have obtained similar
conclusions although they have not taken into account the effects of
rotation.

2. The simulated radial velocities for model C have revealed good
agreement with the data obtained by Bacciotti et al \cite{bacci02},
therefore suggesting that their interpretation that the DG Tau microjet
is rotating must be correct.

3. The inclusion in the models of a helical magnetic field permeating the
jet and the ambient medium, with a maximum value of $\beta \simeq 10$
near the jet surface (corresponding to a toroidal magnetic field $\sim
0.5$ mG), has provided a collimation of the outflow which otherwise
would tend to expand laterally while propagating downstream (Figure 3).
Though the MHD models here investigated were only simple cases involving
rotating continuous (or pulsing) jets $without$ precession,  the results
above suggest that the presence of magnetic fields of the order of
$\sim 0.5$ mG in the DG Tau microjet would be sufficient to collimate
the beam against the centrifugal forces. In fact, a simple estimate
of the magnetic tension forces that would be required to balance the
centrifugal forces at the DG Tau  microjet  surface  gives a  magnetic
field strength $B = (4 \pi  \rho v_{\phi}^2 )^{1/2} \sim 0.2$  mG (where
$\rho \simeq n_{\rm j}  m_{\rm H}\sim  3.3  10^{- 21}$  g cm$^{-3}$ and $v_{\phi}
= 8 $ km s$^{- 1}$), which is compatible with the required value above
from the simulations.  We further notice that this would be the expected
strength of  the magnetic field at radial distances $\sim 40$ AU from the
source. Magnetic flux conservation would then imply strengths $\sim 100$
G near the stellar surface which are compatible both with the values
expected for protostellar sources and also with  magneto-centrifugal
mechanisms for jet production.

\begin{acknowledgements}

We are indebted to Francesca Bacciotti and an anonymous referee for
their very helpful comments and suggestions. They also acknowledge
partial support of the Brazilian agencies FAPESB, FAPESP and CNPq,
the Milenium Institute, and the projects PROPP-UESC (00220.1300.327)
and PRODOC- UFBa (991042-88).

\end{acknowledgements}

\end{document}